\documentclass[11pt,preprint,onecolumn,aps,longbibliography]{revtex4-1}

\usepackage{graphicx}
\usepackage{amsmath} 
\usepackage{mathtools}
\usepackage{amsfonts}
\usepackage{bbm}
\usepackage{braket}
\usepackage{caption}
\usepackage{graphicx}
\usepackage{float}
\usepackage{caption}
\usepackage{subcaption}
\usepackage{url}
\usepackage{color}
\usepackage[ruled]{algorithm2e}

\usepackage{amssymb}
\usepackage{cancel}
\usepackage{upgreek}
\usepackage{tabularx}
\usepackage{color}
\usepackage{units}
\usepackage{hhline}
\usepackage{textfit} 

\usepackage{slashed}
\graphicspath{ {images/} }

\usepackage{graphicx}
\usepackage{dcolumn}
\usepackage{bm}
\usepackage{hyperref}
\usepackage[mathlines]{lineno}

\usepackage{tikz}
\usetikzlibrary{positioning}
\usepackage{qcircuit}
\usepackage{braket}
\usepackage{indentfirst}
\usepackage{float}

\begin{document}


\title{Encoding classical data into a quantum computer}
\author{Kumar J. B. Ghosh}
\email{jb.ghosh@outlook.com}
\affiliation{Department of Electrical and Computer Engineering,
University of Denver, Denver, CO, 80210, USA.
}%





\begin{abstract}
In  this article we describe a technique to transfer data from classical domain to quantum domain. We consider a set of $N (=2^n)$ classical data in the form of a column matrix and prepare a $n$-qubit quantum state, whose components correspond to the $N$ classical data. To prepare this $n$-qubit quantum vector we use Schmidt decomposition and singular value decomposition techniques respectively and construct the corresponding family of quantum circuits. To strengthen our argument we also give  specific examples by considering a set of 4 and 16 classical data and constructing the corresponding  2 and 4-qubit quantum vector respectively.
\end{abstract}

\keywords{quantum computing; Schmidt decomposition; singular value decomposition.}

\maketitle

\newpage

\section{\label{sec:Introduction}Introduction}

Quantum computation is a branch of physics and computer science which solves different computational problems by using quantum-mechanical phenomena, for e.g. entanglement and superposition \cite{Nielsen}. Quantum computation and  quantum information theory is a center of attention in last few decades because it can outperform classical computation and information processing, because the quantum algorithms can give rise to exponential speedups over their classical counterparts. Many known quantum algorithms have a diverse application, such as: integer factorization \cite{Shor}, search algorithm \cite{Grover}, solving constraint satisfection problems \cite{Montanaro}, and quantum machine learning \cite{Mohseni} \cite{Biamonte}. From last few decades three main challanges in quantum computation are: how to encode the classical data into a quantum computer, processing the quantum data, and extracting the useful information from the processed quantum data.  There are few techniques available to tackle the first issue, namely encoding classical data into a quantum state, for e.g. quantum random acess memory (QRAM) proposed by Seth Lloyd et al \cite{Lloyd}, which has distint implications in pattern recognition \cite{Trugenberger}, period finding, quantum fourier transformation etc. A recent article \cite{Cortes}  shows how to classical binary data or bits can be transferred into quantum states. \\

There is another to way to encode classical into quantum computer, namely, to prepare the desired quantum state (described avobe) with the help of controlled rotation in the quantum hilbert space \cite{Barenco}, \cite{Plesch},   \cite{Kumar}. However in this process the number of two-qubit  quantum gates (for e.g. CNOT  gate) required to prepare a general $2n$-qubit state (with real components) is roughly order of $2^{2n}$. A better idea to reduce the number of gates through is described in the follwing article \cite{Bergholm} with the help of Schmidt decomposition  procedure. In this article we describe how to successfully impliment the idea of how to reduce the number of quantum gates (especially two-qubit gates) in the circuit, from the above article  \cite{Bergholm}, with the help of Singular value decomposition procedure \cite{Miszczak}. The reduction in the number of two qubit gates leads to a reduction of complexity in the circuit and also it reduce the error percentage throughout the computation procedure.\\

The outline of this article is as follows. In section \ref{sec:Schmidt decomposition},  we describe the Schmidt decomposition procedure and how to calculate the Schmidt basis and Schmidt coefficients of a quantum vector with the help of singular value decomposition. In   section \ref{sec:Constructing quantum circuit}, we describe how to construct a quantum circuit schematically to  implement this idea.   In section \ref{sec:4qubit example}, we give a specific example by taking a set of $16$ classical data and encode it into a $4$-qubit quantum state. Finally, in section \ref{sec:Discussion and conclusion} we have some discussion of our result and give some concluding remarks. 

\section{\label{sec:Schmidt decomposition}Schmidt decomposition and Singular Value Decomposition}

Before explaining the Schmidt decomposition procedure we first describe an important procedure in linear algebra namely, the singular value decomposition (SVD), which is as follows.\\

Suppose an arbitrary matrix $A\in \mathbb{M}_{m \times n}$ has the rank $k \leq m$. Then there exist the unitary matrices $U\in \mathbb{M}_{m\times m}$ and $V\in \mathbb{M}_{n\times n}$, such that
\begin{equation}
A = U \Sigma V^{\dagger}, \label{svd}
\end{equation}
where the matrix $\Sigma= \lbrace \sigma_{ij}\rbrace \in  \mathbb{M}_{m \times n}$ such that
\begin{equation}
 \sigma_{ij} =0, ~ \text{for}~ i\neq j \label{sigmaii}
\end{equation}
and 
\begin{equation}
 \sigma_{11}\geq \sigma_{22}\geq ... \geq \sigma_{kk}\geq \sigma_{k+1,k+1} =...= \sigma_{qq} =0, \label{sigmaii2}
\end{equation}
with $q= min (m,n)$.\\

We  describe the Schmidt decomposition procedure \cite{Heinosaari},\cite{Bengtsson}  with the help of SVD in the following:

Any pure state $|\psi  \rangle \in \mathbb{C}^{(m+m)}$ can be represented as 
\begin{equation}
|\psi  \rangle = \sum_{i=1}^k \lambda_i |\alpha_i \rangle \otimes |\beta_i \rangle, \label{Schmidt}
\end{equation}
with  $k=\text{min} (m,n)$. The set of bases vectors  $ \lbrace|\alpha_i \rangle \rbrace \in \mathbb{C}^{m}$ and $ \lbrace|\beta_i \rangle\rbrace \in \mathbb{C}^{n}$ are are called Schmidt bases which are orthogonal in respective Hilbert spaces. The coefficients $\lambda_i$ in equation (\ref{Schmidt}) are called Schmidt coefficients.

We can always represent the pure state $|\psi  \rangle \in \mathbb{C}^{(m+m)}$ using computational bases as 
\begin{equation}
|\psi  \rangle =  \sum_{i=1}^m  \sum_{j=1}^n  C_{ij} |e_i \rangle \otimes |f_j \rangle, \label{Schmidt2}
\end{equation}
where $ |e_i \rangle \in \mathbb{C}^{m}$ and $|f_j \rangle\in \mathbb{C}^{n}$. \\

If we do the SVD of the matrix $C$ in (\ref{Schmidt2}) and write 

\begin{equation}
 C = U  \Sigma V^{\dagger}, \label{svd2}
\end{equation}

then without going into the mathematical details, from the three equations (\ref{Schmidt}),  (\ref{Schmidt2}), and  (\ref{svd2}) we can show that 
\begin{equation}
\sigma_{ii} = \lambda_i  , ~ |\alpha_i \rangle = U  |e_i \rangle , ~\text{and}~ |\beta_i \rangle = V^\dagger|f_i \rangle
 \end{equation}
where $\sigma_{ii}$ is described in equations (\ref{sigmaii}) and (\ref{sigmaii2}).\\

Suppose we want to prepare a general $2n$-qubit state. From the method described  in the article \cite{Bergholm} the number of two qubit controlled rotation gates required is order of $2^{2n}$. If we break our general $2n$-qubit state into two $n$-qubit states with the help of Schmidt decomposition and we prepare each of the two $n$-qubit state in  parallel we shall expect  order of $2^{n+1}$ number of controlled rotation gates. This is a significant reduction in the number of two-qubit quantum gates. This argument will be more clear when we describe the method to prepare a general $4$-qubit quantum state.

\section{\label{sec:Constructing quantum circuit}Constructing a schematic circuit for preparing a $\textbf{(2n)}$-$\text{qubit}$ state with Schmidt decomposition}

In this section we give a schematic description to prepare an arbitrary $2n$-qubit state $|\psi  \rangle$ with the help of Schmidt decomposition. In the Schmidt basis the  $2n$-qubit state $|\psi  \rangle$ looks like 
\begin{equation}
|\psi  \rangle =   \sum_{i=1}^N \lambda_i |\alpha_i \rangle \otimes |\beta_i \rangle, \label{schmidt2n}
\end{equation} 
where $N=2^n$. We can calculate  the Schmidt coefficients and corresponding Schmidt bases elements with the help of SVD described in the previous section.\\

We prepare our desired quantum state $|\psi  \rangle$ from initial $|00...0 \rangle$ ($2n$-qubit)   in four steps.

\subsection{\label{subsec:step1}Step 1}

We start from $|00...0  (2n\text{-qubit}) \rangle$. On the first $n$-qubits we apply the rotation gates and construct the $n$-qubit state with $N=2^n$ Schmidt coefficients as follows:
\begin{equation}
|00...0  (2n\text{-qubit}) \rangle \to \left( \sum_{i=1}^N \lambda_i |e_i \rangle \right) |00...0  (n\text{-qubit}) \rangle,
\end{equation}
where $e_i$'s are the basis vectors of the computational basis for $n$-qubits, is defined as:\\
 $$|e_1 \rangle = |00...0\rangle , ~ |e_2 \rangle = |00...01 \rangle ,  ~ |e_3 \rangle = |00...10 \rangle ,~ ...~ |e_N \rangle = |11...11 \rangle. $$

To prepare this state we can use any of  the methods described in \cite{Bergholm} or \cite{Shende} . We expect $N=2^n$ two-qubit quantum gates for this step.

\subsection{\label{subsec:step2}Step 2}

We apply $n$ C-NOT gates from first $n$-qubits to other $n$-qubits respectively. For e.g  first apply C-NOT gate with control on 1st qubit and target on $(n+1)$-th qubit, next apply C-NOT gate with control on 2nd qubit and target on $(n+2)$-th qubit, and so on. This will give rise to the following state:

\begin{equation}
 \left( \sum_{i=1}^N \lambda_i |e_i \rangle \right) |00...0  (n\text{-qubit}) \rangle \to \sum_{i=1}^N \lambda_i |e_i \rangle |e_i \rangle. \label{step2}
\end{equation}

We need $n$ C-NOT gates for this step.

\subsection{\label{subsec:step3}Step 3}

We keep the above form (\ref{step2}) intact and change the $N$ bases states of first $n$-qubits into $N$ bases states  $|\alpha_i \rangle$ described in (\ref{schmidt2n}).

\begin{equation}
|00...0\rangle \to |\alpha_1 \rangle , ~ |00...01 \rangle \to |\alpha_2 \rangle ,  ~ |00...10 \rangle \to |\alpha_3\rangle ,~ ...~ |11...11 \rangle \to |\alpha_N \rangle.
\end{equation}

We expect $N=2^n$ two-qubit quantum gates for this step.

\subsection{\label{subsec:step4}Step 4}

In this step we change the $N$ bases states of remaining $n$-qubits into $N$ bases states  $|\beta_i \rangle$ described in (\ref{schmidt2n}). 
 
\begin{equation}
|00...0\rangle \to |\beta_1 \rangle , ~ |00...01 \rangle \to |\beta_2 \rangle ,  ~ |00...10 \rangle \to |\beta_3 \rangle ,~ ...~ |11...11 \rangle \to |\beta_N \rangle.
\end{equation}

We expect $N=2^n$ two-qubit quantum gates for this step.\\

After the four steps described above we obtain our desired quantum state  $|\psi  \rangle$ in the Schmidt decomposition form. In the  whole four steps the expected number of  two-qubit quantum gates used is order of $(3\times 2^n)$, which is significantly less for large $n$.

\section{\label{sec:4qubit example} Example with $\textbf{2}$-$\text{qubits}$ and $\textbf{4}$-$\text{qubits}$} 

\subsection{{\label{subsec:2qubit}Example with $\textbf{2}$-$\text{qubits}$}}

Let us consider a set of $4$-normalized classical data 

\begin{equation}
X=\left(
\begin{array}{cccc}
 \sqrt{\frac{3}{5}} , & \frac{1}{\sqrt{5}} , &
 \frac{\sqrt{\frac{3}{5}}}{2} , & \frac{1}{2 \sqrt{5}} \\
\end{array}
\right). \label{unnormalizeddata2qubit}
\end{equation}

Form this data our goal is to prepare a normalized two-qubit quantum state $| X \rangle$, where

\begin{equation}
| X \rangle = \sqrt{\frac{3}{5}} | 00 \rangle + \frac{1}{\sqrt{5}} | 01 \rangle+ \frac{1}{2}\sqrt{\frac{3}{5}} |10 \rangle + \frac{1}{2 \sqrt{5}} |11 \rangle.
\end{equation}

We express the quantum data $| X \rangle$ in a compact matrix multiplication form 

\begin{equation}
| X \rangle = e^T C e, \label{matrixform2qubit}
\end{equation}

with superscript $T$ denotes the transpose operation. The matrices $e$ and $C$ are defined as:
\begin{equation}
e = \left(
\begin{array}{c}  
| 0 \rangle \\ | 1 \rangle
\end{array}
\right)~ 
\text{and}~~
C= \left(
\begin{array}{cc}
 \sqrt{\frac{3}{5}} & \frac{1}{\sqrt{5}} \\
 \frac{1}{2}\sqrt{\frac{3}{5}} & \frac{1}{2 \sqrt{5}} \\
\end{array}
\right).\label{matrixCande2qubit}
\end{equation}

If we want to do the Schmidt decomposition of the state $| X \rangle$ we have to do the Singular value decomposition of the matrix $C$ in equation (\ref{matrixCande})

\begin{equation}
 C = U  \Sigma V^{\dagger}. \label{singularvaluedecomposition}
\end{equation} 

We calculate the matrices $ U$,  $\Sigma$, and $V$ as follows:
\begin{equation}
U=\left(
\begin{array}{cc}
 \frac{2}{\sqrt{5}} & -\frac{1}{\sqrt{5}} \\
 \frac{1}{\sqrt{5}} & \frac{2}{\sqrt{5}} \\
\end{array}
\right)~, 
%
~~\Sigma =\left(
\begin{array}{cc}
 1 & 0 \\
 0 & 0 \\
\end{array}
\right), \label{matrixUandSigma2qubit}
\end{equation}
and 
\begin{equation}
V=\left(
\begin{array}{cc}
 \frac{\sqrt{3}}{2} & -\frac{1}{2} \\
 \frac{1}{2} & \frac{\sqrt{3}}{2} \\
\end{array}
\right). \label{matrixV2qubit}
\end{equation}

The diagonal elements of matrix $\Sigma$ in (\ref{matrixV2qubit}) are desired Schmidt coefficients and the matrices $U$ and $V^\dagger$ generates the desired Schmidt bases.\\

If we write the quantum state $| X \rangle$ in a Schmidt decomposition form

\begin{equation}
|X  \rangle = \sum_{i=1}^2 \lambda_i |\alpha_i \rangle \otimes |\beta_i \rangle =  |\alpha_1 \rangle \otimes |\beta_1 \rangle, \label{SchmidtX2qubit}
\end{equation}  

the bases vectors can be obtained from the following equations 

\begin{equation}
\alpha = e^T U ~,~ \text{and} ~ \beta = V^T e, \label{basisstateseg2qubit}
\end{equation}

with the basis vector $e$ has been defined in equation (\ref{matrixCande2qubit}).

Since the matrix $\Sigma$ has only one diagonal element we can easily generate our desired state in the Schmidt decomposed form in the following way.\\

We start with the state $|00 \rangle$. We generate the basis elements $\alpha_1$ and $\beta_1$ by applying rotation gates $U(\theta_1)$ and $U(\theta_2)$ on the two qubits respectively, where the rotation matrix $U(\theta)$ is defined as 

\begin{equation}
U(\theta) = \left(
\begin{array}{cc}
 \cos\theta & -\sin\theta  \\
 sin \theta & \cos \theta \\
\end{array}
\right). \label{utheta}
\end{equation}  

In our case we can find out $\theta_1 = \arcsin(1/\sqrt{5})$ and $\theta_2 = \arcsin(1/2)$ respectively. 

\subsection{{\label{subsec:4qubit}Example with $\textbf{4}$-$\text{qubits}$}}

Let us consider the following set of $16$ un-normalized classical data: 
\begin{equation}
X_{unnormal}=\left\{3 \sqrt{3},-\sqrt{3},-9,3,-2 \sqrt{3},-6 \sqrt{3},6,18,-3 \sqrt{3},\sqrt{3},-3,1,2 \sqrt{3},6 \sqrt{3},2,6\right\} \label{unnormalizeddata}
\end{equation}

Form this data our goal is to prepare a normalized quantum state $| X \rangle$ where

\begin{eqnarray}
| X \rangle &=& \frac{3 \sqrt{\frac{3}{2}}}{20}| 0000 \rangle  -\frac{\sqrt{\frac{3}{2}}}{20}| 0001 \rangle  -\frac{9}{20 \sqrt{2}}| 0010 \rangle +\frac{3}{20 \sqrt{2}}| 0011 \rangle  -\frac{\sqrt{\frac{3}{2}}}{10}| 0100 \rangle  -\frac{3\sqrt{\frac{3}{2}}}{10}| 0101 \rangle  \nonumber \\ 
&+& \frac{3}{10 \sqrt{2}}| 0110 \rangle +\frac{9}{10 \sqrt{2}}| 0111 \rangle -\frac{3 \sqrt{\frac{3}{2}}}{20}| 1000 \rangle +\frac{\sqrt{\frac{3}{2}}}{20} | 1001 \rangle  -\frac{3}{20 \sqrt{2}} | 1010 \rangle +\frac{1}{20
   \sqrt{2}} | 1011 \rangle  \nonumber \\ 
   &+& \frac{\sqrt{\frac{3}{2}}}{10} | 1100 \rangle +\frac{3 \sqrt{\frac{3}{2}}}{10} | 1101 \rangle +\frac{1}{10 \sqrt{2}} | 1110 \rangle +\frac{3}{10 \sqrt{2}}   | 1111 \rangle. \label{normalizedvector}
\end{eqnarray}

We  express our quantum data $| X \rangle$ in a compact matrix multiplication form 

\begin{equation}
| X \rangle = e^T C e, \label{matrixform}
\end{equation}

with superscript $T$ denotes the transpose operation. The matrices $e$ and $C$ are defined as

\begin{equation}
e = \left(
\begin{array}{c}  
| 00 \rangle \\ | 10 \rangle \\ | 01 \rangle \\ | 11 \rangle \\
\end{array}
\right)~ 
\text{and}~~
C= \left(
\begin{array}{cccc}
 \frac{3 \sqrt{\frac{3}{2}}}{20} & -\frac{\sqrt{\frac{3}{2}}}{20} & -\frac{9}{20 \sqrt{2}} & \frac{3}{20 \sqrt{2}} \\
 -\frac{\sqrt{\frac{3}{2}}}{10} & -\frac{3 \sqrt{\frac{3}{2}}}{10} & \frac{3}{10 \sqrt{2}} & \frac{9}{10 \sqrt{2}} \\
 -\frac{3 \sqrt{\frac{3}{2}}}{20} & \frac{\sqrt{\frac{3}{2}}}{20} & -\frac{3}{20 \sqrt{2}} & \frac{1}{20 \sqrt{2}} \\
 \frac{\sqrt{\frac{3}{2}}}{10} & \frac{3 \sqrt{\frac{3}{2}}}{10} & \frac{1}{10 \sqrt{2}} & \frac{3}{10 \sqrt{2}} \\
\end{array}
\right).\label{matrixCande}
\end{equation}

If we want to do the Schmidt decomposition of the state $| X \rangle$ we have to do the Singular value decomposition of the matrix $C$ in equation (\ref{matrixCande})

\begin{equation}
 C = U  \Sigma V^{\dagger}. \label{singularvaluedecomposition}
\end{equation} 

We calculate the matrices $ U$,  $\Sigma$, and $V$ 

\begin{equation}
U=\left(
\begin{array}{cccc}
 0 & 0 & 1 & 0 \\
 1 & 0 & 0 & 0 \\
 0 & 0 & 0 & 1 \\
 0 & 1 & 0 & 0 \\
\end{array}
\right)~, 
%
~~\Sigma =\left(
\begin{array}{cccc}
 \sqrt{\frac{3}{5}} & 0 & 0 & 0 \\
 0 & \frac{1}{\sqrt{5}} & 0 & 0 \\
 0 & 0 & \frac{\sqrt{\frac{3}{5}}}{2} & 0 \\
 0 & 0 & 0 & \frac{1}{2 \sqrt{5}} \\
\end{array}
\right), \label{matrixUandSigma}
\end{equation}
and 
\begin{equation}
V=\left(
\begin{array}{cccc}
 -\frac{1}{2 \sqrt{10}} & \frac{\sqrt{\frac{3}{10}}}{2} & \frac{3}{2 \sqrt{10}} & -\frac{3 \sqrt{\frac{3}{10}}}{2} \\
 -\frac{3}{2 \sqrt{10}} & \frac{3 \sqrt{\frac{3}{10}}}{2} & -\frac{1}{2 \sqrt{10}} & \frac{\sqrt{\frac{3}{10}}}{2} \\
 \frac{\sqrt{\frac{3}{10}}}{2} & \frac{1}{2 \sqrt{10}} & -\frac{3 \sqrt{\frac{3}{10}}}{2} & -\frac{3}{2 \sqrt{10}} \\
 \frac{3 \sqrt{\frac{3}{10}}}{2} & \frac{3}{2 \sqrt{10}} & \frac{\sqrt{\frac{3}{10}}}{2} & \frac{1}{2 \sqrt{10}} \\
\end{array}
\right). \label{matrixV}
\end{equation}

The diagonal elements of matrix $\Sigma$ in (\ref{matrixV}) are desired Schmidt coefficients and the matrices $U$ and $V^\dagger$ generates the desired Schmidt bases.\\

If we write the quantum state $| X \rangle$ in a Schmidt decomposition form
\begin{equation}
|X  \rangle = \sum_{i=1}^4 \lambda_i |\alpha_i \rangle \otimes |\beta_i \rangle, \label{SchmidtX}
\end{equation}
  
the bases vectors can be obtained from the following equations 

\begin{equation}
\alpha = e^T U ~,~ \text{and} ~ \beta = V^T e, \label{basisstateseg}
\end{equation}

with 
$$e = \left(
\begin{array}{cccc}  
| 00 \rangle , & | 10 \rangle , & | 01 \rangle , & | 11 \rangle 
\end{array}
\right)^T$$ 
are the computational basis vectors defined in equation (\ref{matrixCande}). In our case we can construct the quantum circuit in the following four steps:

\subsubsection{\label{subsubsec:step1}Step 1}

We start from $|0000 \rangle$. On the first $2$-qubits we apply two rotation gates and construct the $2$-qubit state with $4$ Schmidt coefficients as follows:
\begin{equation}
|0000 \rangle \to \left( \sum_{i=1}^4 \lambda_i |e_i \rangle \right) |00 \rangle,
\end{equation}
with 
\begin{equation}
|e_1 \rangle = |00 \rangle , ~ |e_2 \rangle= |01 \rangle, ~|e_3 \rangle = |10 \rangle , ~\text{and}~|e_4 \rangle = |11 \rangle.
\end{equation}

To prepare this state we apply the rotation gate $U(\theta)$ on the first two qubits respectively, where the rotation matrix $U(\theta)$ is defined as

\begin{equation}
U(\theta) = \left(
\begin{array}{cc}
 \cos\theta & -\sin\theta  \\
 sin \theta & \cos \theta \\
\end{array}
\right). \label{utheta}
\end{equation}

For the first qubit $\theta_1 = \arcsin(1/2)$, and for the second qubit $\theta_2 = \arcsin(2/\sqrt{5})$ will give rise to the four Schmidt coefficients $\sqrt{\frac{3}{5}}, \frac{1}{\sqrt{5}}, \frac{\sqrt{3}}{2\sqrt{5}}$, and $\frac{1}{2 \sqrt{5}}$ respectively.

\subsubsection{\label{subsubsec:step2}Step 2}

We apply two C-NOT gates respectively: one with control on 1st qubit and target on 3rd qubit,   another with control on 2nd qubit and target on 4th qubit. This will give rise to the following state

\begin{equation}
 \left( \sum_{i=1}^4 \lambda_i |e_i \rangle \right) |00 \rangle \to \sum_{i=1}^4 \lambda_i |e_i \rangle |e_i \rangle. \label{step21}
\end{equation}

We need $2$ C-NOT gates for this step.

\subsubsection{\label{subsubsec:step3}Step 3}

We keep the above form (\ref{step21}) intact and change the $4$ bases states of first $2$-qubits into $4$ bases states  $|\alpha_i \rangle$ described in (\ref{basisstateseg}), which is as follows:
\begin{equation}
|00\rangle \to |\alpha_1 \rangle = |01\rangle , ~ |01 \rangle \to |\alpha_2 \rangle=|11\rangle ,  ~ |10 \rangle \to |\alpha_3\rangle=|00\rangle ,~ |11 \rangle \to |\alpha_4 \rangle=|10\rangle.
\end{equation}

To do this transformation we first apply a Pauli-X gate on the first qubit and then we put a SWAP gate on the first two qubits. 

\subsubsection{\label{subsubsec:step4}Step 4}

In this step we change the $4$ bases states of remaining $2$-qubits (third and fourth) into $4$ bases states  $|\beta_i \rangle$ described in (\ref{schmidt2n})
 
\begin{equation}
|00\rangle \to |\beta_1 \rangle , ~ |01 \rangle \to |\beta_2 \rangle ,  ~ |10 \rangle \to |\beta_3 \rangle ,~ \text{and}~ |11 \rangle \to |\beta_4 \rangle,
\end{equation}

where $\beta = V^T e $, with the matrix $V$ is defined in equation (\ref{matrixV}).\\

To prepare this state we apply the rotation gate $U(\theta)$ on the last two qubits respectively, where the rotation matrix $U(\theta)$ is defined in equation (\ref{utheta}). For the third qubit $\theta_3 = \arcsin(\sqrt{3}/2)$, and for the fourth qubit $\theta_4 = \arcsin(1/\sqrt{10})$ will give rise to the transformation matrix $V^T$. \\

Though it is very difficult to construct the quantum circuit which gives rise to a general basis transformation matrix but we can always find a suitable quantum circuit where the number of two-qubit gates will not exceed $2^n$ where $n$ is the number of corresponding qubits. For more details please look into the articles  \cite{Vatan} \cite{Vidal}, and \cite{li2013decomposition}.

\subsubsection{\label{subsubsec:Sample quantum circuit}Sample quantum circuit} 
In Fig. we describe quantum circuit for the example we have described above:

\begin{figure}[H]
    \centering
    \begin{tikzpicture}[
 image/.style = {text width=0.7\textwidth, 
                 inner sep=0pt, outer sep=0pt},
node distance = 1mm and 1mm
                        ] 
\node [image] (frame1)
    {\includegraphics[width=\linewidth]{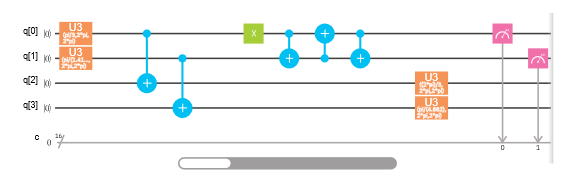}};
\end{tikzpicture}
\caption{Quantum circuit for encoding the classical data described above.}
\label{energy compare bcc sc}
\end{figure} 

Though this circuit varies with the values of our classical data.

\section{\label{sec:Discussion and conclusion}Discussion and conclusion}

In this section we shall make few comments about our result.\\

Though we have considered even number of qubits in most our calculation but we can generalize our scheme to odd number of qubits $(2n+1)$ also. For this we divide our initial $(2n+1)$ qubits into two sets:  $(n)$-qubits and $(n+1)$-qubits respectively. After that we construct the corresponding matrix $C$ described in (\ref{svd2} ,  \ref{matrixCande}) and do the singular value decomposition to obtain the corresponding unitary matrices $ U$,  $\Sigma$, and $V$ and find the corresponding Schmidt coefficients and Schmidt bases.\\

The challenges of this method are: calculating SVD which takes good classical computing resources and finding a quantum circuit for the basis transformation at the end. Since basis transformation matrices are unitary so we can always find a series of quantum gates to perform this operation, but to find the right combination of quantum gates is tedious work. But if we can spare the classical computation part, and construct our desired quantum vector,  we can use it to compute more complicated algorithms which will take enormous amount of time to compute in the classical computers.

\bibliography{Classical_to_quantum}

\begin{thebibliography}{20}%
\makeatletter
\providecommand \@ifxundefined [1]{%
 \@ifx{#1\undefined}
}%
\providecommand \@ifnum [1]{%
 \ifnum #1\expandafter \@firstoftwo
 \else \expandafter \@secondoftwo
 \fi
}%
\providecommand \@ifx [1]{%
 \ifx #1\expandafter \@firstoftwo
 \else \expandafter \@secondoftwo
 \fi
}%
\providecommand \natexlab [1]{#1}%
\providecommand \enquote  [1]{``#1''}%
\providecommand \bibnamefont  [1]{#1}%
\providecommand \bibfnamefont [1]{#1}%
\providecommand \citenamefont [1]{#1}%
\providecommand \href@noop [0]{\@secondoftwo}%
\providecommand \href [0]{\begingroup \@sanitize@url \@href}%
\providecommand \@href[1]{\@@startlink{#1}\@@href}%
\providecommand \@@href[1]{\endgroup#1\@@endlink}%
\providecommand \@sanitize@url [0]{\catcode `\\12\catcode `\$12\catcode
  `\&12\catcode `\#12\catcode `\^12\catcode `\_12\catcode `\%12\relax}%
\providecommand \@@startlink[1]{}%
\providecommand \@@endlink[0]{}%
\providecommand \url  [0]{\begingroup\@sanitize@url \@url }%
\providecommand \@url [1]{\endgroup\@href {#1}{\urlprefix }}%
\providecommand \urlprefix  [0]{URL }%
\providecommand \Eprint [0]{\href }%
\providecommand \doibase [0]{http://dx.doi.org/}%
\providecommand \selectlanguage [0]{\@gobble}%
\providecommand \bibinfo  [0]{\@secondoftwo}%
\providecommand \bibfield  [0]{\@secondoftwo}%
\providecommand \translation [1]{[#1]}%
\providecommand \BibitemOpen [0]{}%
\providecommand \bibitemStop [0]{}%
\providecommand \bibitemNoStop [0]{.\EOS\space}%
\providecommand \EOS [0]{\spacefactor3000\relax}%
\providecommand \BibitemShut  [1]{\csname bibitem#1\endcsname}%
\let\auto@bib@innerbib\@empty
\bibitem [{\citenamefont {Nielsen}\ and\ \citenamefont {L.}(2001)}]{Nielsen}%
  \BibitemOpen
  \bibfield  {author} {\bibinfo {author} {\bibfnamefont {M.~A.}\ \bibnamefont
  {Nielsen}}\ and\ \bibinfo {author} {\bibfnamefont {Chuang~I.}\ \bibnamefont
  {L.}},\ }\href@noop {} {\emph {\bibinfo {title} {Quantum Computation and
  Quantum Information}}}\ (\bibinfo  {publisher} {Cambridge University Press},\
  \bibinfo {address} {Cambridge},\ \bibinfo {year} {2001})\BibitemShut
  {NoStop}%
\bibitem [{\citenamefont {Shor}(1994)}]{Shor}%
  \BibitemOpen
  \bibfield  {author} {\bibinfo {author} {\bibfnamefont {P.~W.}\ \bibnamefont
  {Shor}},\ }\bibfield  {title} {\enquote {\bibinfo {title} {Algorithms for
  quantum computation: discrete logarithms and factoring},}\ }\href@noop {}
  {\bibfield  {journal} {\bibinfo  {journal} {Proceedings 35th annual symposium
  on foundations of computer science}\ ,\ \bibinfo {pages} {124--134}}
  (\bibinfo {year} {1994})}\BibitemShut {NoStop}%
\bibitem [{\citenamefont {Grover}(1997)}]{Grover}%
  \BibitemOpen
  \bibfield  {author} {\bibinfo {author} {\bibfnamefont {L.~K.}\ \bibnamefont
  {Grover}},\ }\bibfield  {title} {\enquote {\bibinfo {title} {Quantum
  mechanics helps in searching for a needle in a haystack},}\ }\href@noop {}
  {\bibfield  {journal} {\bibinfo  {journal} {Phys. Rev. Lett.}\ }\textbf
  {\bibinfo {volume} {79}},\ \bibinfo {pages} {325} (\bibinfo {year}
  {1997})}\BibitemShut {NoStop}%
\bibitem [{\citenamefont {Montanaro}(2015)}]{Montanaro}%
  \BibitemOpen
  \bibfield  {author} {\bibinfo {author} {\bibfnamefont {A.}~\bibnamefont
  {Montanaro}},\ }\href@noop {} {\enquote {\bibinfo {title} {Quantum walk
  speedup of backtracking algorithms},}\ } (\bibinfo {year} {2015})\BibitemShut
  {NoStop}%
\bibitem [{\citenamefont {Lloyd}\ \emph {et~al.}(2013)\citenamefont {Lloyd},
  \citenamefont {Mohseni},\ and\ \citenamefont {Rebentrost}}]{Mohseni}%
  \BibitemOpen
  \bibfield  {author} {\bibinfo {author} {\bibfnamefont {S.}~\bibnamefont
  {Lloyd}}, \bibinfo {author} {\bibfnamefont {M.}~\bibnamefont {Mohseni}}, \
  and\ \bibinfo {author} {\bibfnamefont {P.}~\bibnamefont {Rebentrost}},\
  }\href@noop {} {\enquote {\bibinfo {title} {Quantum algorithms for supervised
  and unsupervised machine learning},}\ } (\bibinfo {year} {2013})\BibitemShut
  {NoStop}%
\bibitem [{\citenamefont {Biamonte}(2017)}]{Biamonte}%
  \BibitemOpen
  \bibfield  {author} {\bibinfo {author} {\bibfnamefont {Jacob et~al.}\
  \bibnamefont {Biamonte}},\ }\bibfield  {title} {\enquote {\bibinfo {title}
  {Quantum machine learning},}\ }\href@noop {} {\bibfield  {journal} {\bibinfo
  {journal} {Nature}\ }\textbf {\bibinfo {volume} {549}},\ \bibinfo {pages}
  {195--202} (\bibinfo {year} {2017})}\BibitemShut {NoStop}%
\bibitem [{\citenamefont {Giovannetti}\ \emph {et~al.}(2008)\citenamefont
  {Giovannetti}, \citenamefont {Lloyd},\ and\ \citenamefont {Maccone}}]{Lloyd}%
  \BibitemOpen
  \bibfield  {author} {\bibinfo {author} {\bibfnamefont {V.}~\bibnamefont
  {Giovannetti}}, \bibinfo {author} {\bibfnamefont {S.}~\bibnamefont {Lloyd}},
  \ and\ \bibinfo {author} {\bibfnamefont {L.}~\bibnamefont {Maccone}},\
  }\bibfield  {title} {\enquote {\bibinfo {title} {Quantum random access
  memory},}\ }\href@noop {} {\bibfield  {journal} {\bibinfo  {journal} {Phys.
  Rev. Lett.}\ }\textbf {\bibinfo {volume} {100}},\ \bibinfo {pages} {160501}
  (\bibinfo {year} {2008})}\BibitemShut {NoStop}%
\bibitem [{\citenamefont {Trugenberger}(2001)}]{Trugenberger}%
  \BibitemOpen
  \bibfield  {author} {\bibinfo {author} {\bibfnamefont {C.~A.}\ \bibnamefont
  {Trugenberger}},\ }\bibfield  {title} {\enquote {\bibinfo {title}
  {Probabilistic quantum memories},}\ }\href@noop {} {\bibfield  {journal}
  {\bibinfo  {journal} {Phys. Rev. Lett.}\ }\textbf {\bibinfo {volume} {87}},\
  \bibinfo {pages} {067901} (\bibinfo {year} {2001})}\BibitemShut {NoStop}%
\bibitem [{\citenamefont {Cortese}\ and\ \citenamefont {Braje}(2018)}]{Cortes}%
  \BibitemOpen
  \bibfield  {author} {\bibinfo {author} {\bibfnamefont {J.~A.}\ \bibnamefont
  {Cortese}}\ and\ \bibinfo {author} {\bibfnamefont {T.~M.}\ \bibnamefont
  {Braje}},\ }\href@noop {} {\enquote {\bibinfo {title} {Loading classical data
  into a quantum computer},}\ } (\bibinfo {year} {2018})\BibitemShut {NoStop}%
\bibitem [{\citenamefont {Barenco}\ and\ \citenamefont
  {et~al.}(1995)}]{Barenco}%
  \BibitemOpen
  \bibfield  {author} {\bibinfo {author} {\bibfnamefont {A.}~\bibnamefont
  {Barenco}}\ and\ \bibinfo {author} {\bibfnamefont {Bennett}\ \bibnamefont
  {et~al.}},\ }\bibfield  {title} {\enquote {\bibinfo {title} {Elementary gates
  for quantum computation},}\ }\href@noop {} {\bibfield  {journal} {\bibinfo
  {journal} {Phys. Rev. A}\ }\textbf {\bibinfo {volume} {52}},\ \bibinfo
  {pages} {3457} (\bibinfo {year} {1995})}\BibitemShut {NoStop}%
\bibitem [{\citenamefont {Plesch}\ and\ \citenamefont
  {Brukner}(2011)}]{Plesch}%
  \BibitemOpen
  \bibfield  {author} {\bibinfo {author} {\bibfnamefont {M.}~\bibnamefont
  {Plesch}}\ and\ \bibinfo {author} {\bibfnamefont {{\v{C}}aslav}\ \bibnamefont
  {Brukner}},\ }\bibfield  {title} {\enquote {\bibinfo {title} {Quantum-state
  preparation with universal gate decompositions},}\ }\href@noop {} {\bibfield
  {journal} {\bibinfo  {journal} {Phys. Rev. A}\ }\textbf {\bibinfo {volume}
  {83}},\ \bibinfo {pages} {032302} (\bibinfo {year} {2011})}\BibitemShut
  {NoStop}%
\bibitem [{\citenamefont {Kumar}(2013)}]{Kumar}%
  \BibitemOpen
  \bibfield  {author} {\bibinfo {author} {\bibfnamefont {P.}~\bibnamefont
  {Kumar}},\ }\bibfield  {title} {\enquote {\bibinfo {title} {Direct
  implementation of an n-qubit controlled-unitary gate in a single step},}\
  }\href@noop {} {\bibfield  {journal} {\bibinfo  {journal} {Quantum
  information processing}\ }\textbf {\bibinfo {volume} {12}},\ \bibinfo {pages}
  {1201--1223} (\bibinfo {year} {2013})}\BibitemShut {NoStop}%
\bibitem [{\citenamefont {Bergholm}\ \emph {et~al.}(2005)\citenamefont
  {Bergholm}, \citenamefont {Vartiainen}, \citenamefont {M{\"o}tt{\"o}nen},\
  and\ \citenamefont {Salomaa}}]{Bergholm}%
  \BibitemOpen
  \bibfield  {author} {\bibinfo {author} {\bibfnamefont {V.}~\bibnamefont
  {Bergholm}}, \bibinfo {author} {\bibfnamefont {J.~J.}\ \bibnamefont
  {Vartiainen}}, \bibinfo {author} {\bibfnamefont {M.}~\bibnamefont
  {M{\"o}tt{\"o}nen}}, \ and\ \bibinfo {author} {\bibfnamefont {M.~M.}\
  \bibnamefont {Salomaa}},\ }\bibfield  {title} {\enquote {\bibinfo {title}
  {Quantum circuits with uniformly controlled one-qubit gates},}\ }\href@noop
  {} {\bibfield  {journal} {\bibinfo  {journal} {Phys. Rev. A}\ }\textbf
  {\bibinfo {volume} {71}},\ \bibinfo {pages} {052330} (\bibinfo {year}
  {2005})}\BibitemShut {NoStop}%
\bibitem [{\citenamefont {Miszczak}(2011)}]{Miszczak}%
  \BibitemOpen
  \bibfield  {author} {\bibinfo {author} {\bibfnamefont {J.~A.}\ \bibnamefont
  {Miszczak}},\ }\bibfield  {title} {\enquote {\bibinfo {title} {Singular value
  decomposition and matrix reorderings in quantum information theory},}\
  }\href@noop {} {\bibfield  {journal} {\bibinfo  {journal} {International
  Journal of Modern Physics C}\ }\textbf {\bibinfo {volume} {22}},\ \bibinfo
  {pages} {897--918} (\bibinfo {year} {2011})}\BibitemShut {NoStop}%
\bibitem [{\citenamefont {Heinosaari}\ and\ \citenamefont
  {Ziman}(2008)}]{Heinosaari}%
  \BibitemOpen
  \bibfield  {author} {\bibinfo {author} {\bibfnamefont {T.}~\bibnamefont
  {Heinosaari}}\ and\ \bibinfo {author} {\bibfnamefont {M.}~\bibnamefont
  {Ziman}},\ }\bibfield  {title} {\enquote {\bibinfo {title} {Guide to
  mathematical concepts of quantum theory},}\ }\href@noop {} {\bibfield
  {journal} {\bibinfo  {journal} {AcPSl}\ }\textbf {\bibinfo {volume} {58}},\
  \bibinfo {pages} {487--674} (\bibinfo {year} {2008})}\BibitemShut {NoStop}%
\bibitem [{\citenamefont {Bengtsson}\ and\ \citenamefont
  {Zyczkowski}(2006)}]{Bengtsson}%
  \BibitemOpen
  \bibfield  {author} {\bibinfo {author} {\bibfnamefont {I.}~\bibnamefont
  {Bengtsson}}\ and\ \bibinfo {author} {\bibfnamefont {K.}~\bibnamefont
  {Zyczkowski}},\ }\href@noop {} {\enquote {\bibinfo {title} {Geometry of
  quantum states cambridge university press},}\ } (\bibinfo {year}
  {2006})\BibitemShut {NoStop}%
\bibitem [{\citenamefont {Shende}\ \emph {et~al.}(2006)\citenamefont {Shende},
  \citenamefont {Bullock},\ and\ \citenamefont {Markov}}]{Shende}%
  \BibitemOpen
  \bibfield  {author} {\bibinfo {author} {\bibfnamefont {V.~V.}\ \bibnamefont
  {Shende}}, \bibinfo {author} {\bibfnamefont {S.~S.}\ \bibnamefont {Bullock}},
  \ and\ \bibinfo {author} {\bibfnamefont {I.~L.}\ \bibnamefont {Markov}},\
  }\bibfield  {title} {\enquote {\bibinfo {title} {Synthesis of quantum-logic
  circuits},}\ }\href@noop {} {\bibfield  {journal} {\bibinfo  {journal} {IEEE
  Transactions on Computer-Aided Design of Integrated Circuits and Systems}\
  }\textbf {\bibinfo {volume} {25}},\ \bibinfo {pages} {1000--1010} (\bibinfo
  {year} {2006})}\BibitemShut {NoStop}%
\bibitem [{\citenamefont {Vatan}\ and\ \citenamefont {Williams}(2004)}]{Vatan}%
  \BibitemOpen
  \bibfield  {author} {\bibinfo {author} {\bibfnamefont {F.}~\bibnamefont
  {Vatan}}\ and\ \bibinfo {author} {\bibfnamefont {C.}~\bibnamefont
  {Williams}},\ }\bibfield  {title} {\enquote {\bibinfo {title} {Optimal
  quantum circuits for general two-qubit gates},}\ }\href@noop {} {\bibfield
  {journal} {\bibinfo  {journal} {Phys. Rev. A}\ }\textbf {\bibinfo {volume}
  {69}},\ \bibinfo {pages} {032315} (\bibinfo {year} {2004})}\BibitemShut
  {NoStop}%
\bibitem [{\citenamefont {Vidal}\ and\ \citenamefont {Dawson}(2004)}]{Vidal}%
  \BibitemOpen
  \bibfield  {author} {\bibinfo {author} {\bibfnamefont {G.}~\bibnamefont
  {Vidal}}\ and\ \bibinfo {author} {\bibfnamefont {C.~M.}\ \bibnamefont
  {Dawson}},\ }\bibfield  {title} {\enquote {\bibinfo {title} {Universal
  quantum circuit for two-qubit transformations with three controlled-not
  gates},}\ }\href@noop {} {\bibfield  {journal} {\bibinfo  {journal} {Phys.
  Rev. A}\ }\textbf {\bibinfo {volume} {69}},\ \bibinfo {pages} {010301}
  (\bibinfo {year} {2004})}\BibitemShut {NoStop}%
\bibitem [{\citenamefont {Li}\ \emph {et~al.}(2013)\citenamefont {Li},
  \citenamefont {Roberts},\ and\ \citenamefont {Yin}}]{li2013decomposition}%
  \BibitemOpen
  \bibfield  {author} {\bibinfo {author} {\bibfnamefont {C.~K.}\ \bibnamefont
  {Li}}, \bibinfo {author} {\bibfnamefont {R.}~\bibnamefont {Roberts}}, \ and\
  \bibinfo {author} {\bibfnamefont {X.}~\bibnamefont {Yin}},\ }\bibfield
  {title} {\enquote {\bibinfo {title} {Decomposition of unitary matrices and
  quantum gates},}\ }\href@noop {} {\bibfield  {journal} {\bibinfo  {journal}
  {International Journal of Quantum Information}\ }\textbf {\bibinfo {volume}
  {11}},\ \bibinfo {pages} {1350015} (\bibinfo {year} {2013})}\BibitemShut
  {NoStop}%
\end{thebibliography}%

\end{document}